# CONTRIBUTION OF SOCIAL NETWORK ANALYSIS AND COLLECTIVE PHENOMENA TO UNDERSTANDING SOCIAL COMPLEXITY AND COGNITION

Denis Boyer[1,2]

Gabriel Ramos-Fernandez[2,3]

[1] Instituto de Física, Universidad Nacional Autónoma de México, Ciudad de México 04510, México.

[2] Centro de Ciencias de la Complejidad, Universidad Nacional Autónoma de México, Ciudad de México 04510 México.

[3] Centro Interdisciplinario de Investigación para el Desarrollo Integral Regional (CIIDIR) Unidad Oaxaca, Instituto Politécnico Nacional, Hornos 1003, Santa Cruz Xoxocotlán, Oaxaca 71230, México.

Denis Boyer: boyer@fisica.unam.mx ; Gabriel Ramos-Fernandez: ramosfer@alumni.upenn.edu

**Abstract**

The social brain hypothesis postulates the increasing complexity of social interactions as a driving force for the evolution of cognitive abilities. Whereas dyadic and triadic relations play a basic role in defining social behaviours and pose many challenges for the social brain, individuals in animal societies typically belong to relatively large networks. How the structure and dynamics of these networks also contribute to the evolution of cognition, and vice versa, is less understood. Here we review how collective phenomena can occur in systems where social agents do not require sophisticated cognitive skills, and how complex networks can grow from simple probabilistic rules, or even emerge from the interaction between agents and their environment, without explicit social factors. We further show that the analysis of social networks can be used to develop good indicators of social complexity beyond the individual or dyadic level. We also discuss the types of challenges that the social brain must cope with in structured groups, such as higher information fluxes, originating from individuals playing different roles in the network, or dyadic contacts of widely varying durations and frequencies. We discuss the relevance of these ideas for primates and other animals' societies.

# 1. Introduction

The social brain, social complexity or social intelligence hypothesis (SCH; Jolly 1966; Humphrey 1976; Dunbar 1988) postulates that selective pressures originating from the social environment played a major role in the evolution of cognitive abilities of group-living species, particularly in primates. This hypothesis assumes that individuals in complex societies must be able to keep track of moment-to-moment interactions with different individuals, integrate these social interactions into general patterns of social relationships and use this knowledge to predict others' behaviour. Because these abilities are cognitively challenging, it is supposed that an increase in social complexity necessarily entails an increase in cognitive complexity.

The SCH has been criticised for being anthropomorphic (Barrett et al. 2007), attributing human-like cognitive features to other animals without considering alternative ways of dealing with a complex social environment. In addition, there has been some debate on what actually constitutes social complexity (Hemelrijk 2002; Barrett et al. 2007; Barrett et al. 2012; Bergman and Beehner 2015). In this chapter we review how models of collective behaviour can provide insights into the cognitive abilities that are minimally required for social structures to emerge out of simple interaction patterns. We also review how analysing social networks using formal metrics can provide useful measures of social complexity which may help to ground the discussion on what is actually challenging about the social environment.

## 2. What is social complexity?

Traditionally, what has been considered challenging about social behaviour in primates is the existence of triadic relationships: it is not only important that two individuals know about each other's behaviour, but also how would each behave in the presence of a third individual (Cheney and Seyfarth 1990). The possibility of coalition formation has been noted as one of the hallmarks of social complexity in primates (Silk et al. 2004) and carnivores (Smith et al. 2010). Proponents of the SCH argue that in order to predict whether another individual will participate in a coalition or not, individuals must use abstractions such as dominance rank or matrilineal relationship. At the core of the debate about whether such abstractions are necessary or not is the issue of which higher-level properties influence lower levels of social organisation: does the dominance hierarchy, as the general pattern of social relationships, influence moment-to-moment interactions? More crucially, do individuals use these general patterns to predict how will others behave? In the case of dominance relationships, is the concept of dominance rank necessary for making decisions by the individuals involved in those interactions?

We argue that theories that focus only on interactions and relationships, even if useful, cannot explain all the complexity of social behaviour in primates. It is tempting to relate social complexity with the number of interactions an individual can process during a short period of time. However, such framework may be difficult to scale up. Exploring higher levels of complexity beyond triadic relationships seems complicated and impractical, since it would require to take into account tetradic, pentadic interactions (and so forth). The presence of many, practically simultaneous stimuli is likely to complicate decision making, without necessarily implying the recognition of new social patterns. A different approach, developed throughout this chapter, consists in determining how a small number of relatively simple interactions between neighbouring individuals can self-organise and generate macroscopic structures at the group level and how this can inform us about the actual complexity faced by individual group members.

Flack (2012) proposes that in order to understand social complexity we should specify how macroscopic or aggregate properties (such as social structure) arise from microscopic dynamics (such as temporally dynamic social interactions). She suggests that higher organisational levels such as the social structure arise when coarse-grained representations of social interactions at lower levels become useful for animals to make decisions. This paradigm is highly relevant for social cognition: if individual behaviour is indeed influenced by higher levels of organisation (so called "slow variables" that change more slowly than day-to-day interactions), it means that individuals should have the computational capacity to consider them. In other words, any component of the system, if it is to be influenced by the statistical regularities of the system's behaviour, should be able to record and process these regularities in order to make decisions about its behaviour.

Barrett et al (2007) suggest that social complexity theory could benefit from an embodied approach, where cognition is the engagement with the world in terms of perception and action mechanisms, not only in terms of abstract mental constructs (like dominance rank in the example above). This engagement allows individuals to exploit the structure of the environment. The discovery of mirror neurons (Di Pellegrino et al. 1992), which activate when performing an action but also when others perform it, suggests that monkeys could have an automatic (and likely unconscious) understanding of others as goal-directed agents (Gallese & Goldman1998). This could actually constitute a form of pattern recognition involving active perception, which implies a social understanding but not requiring social concepts. In fact, this pattern recognition could be what individuals, according to Flack (2012) need to respond to the statistical regularities of the slow variables. Pattern recognition of social features could also be updated as the social environment changes, using the most familiar configuration to act on the world. This experientially informed pattern recognition could be expensive in terms of neuronal tissue and connectivity (which would give support to the social brain hypothesis as an explanation for large brains).

## 3. Emerging complexity in collective phenomena

Some aspects of social behaviour can be complex for collective reasons rather than because of the sophistication of the cognitive tasks to be performed by individuals. Viewing groups as systems of interacting elements, tools borrowed from physics can be useful to elaborate simple models that help to understand how other forms of complexity, not reduced to that of the individual brain, can emerge in social groups. We focus on two classes of problems, which are often inter-related. One is emergence itself, namely, the birth of collective patterns which cannot be predicted from the study of individuals or their pair-wise (or triadic) interactions taken in isolation. The second class aims to understand the functions of collective structures, such as networks. We will discuss their stability and resilience, and more generally, how collective aspects can orient individual behaviour itself and give rise to certain dynamics in response to internal or external changes.

**3.1 From small-scale interactions to macroscopic patterns in systems of agents**

A distinctive feature of collective patterns is that they exhibit some kind of order, for instance coordination, that can emerge without a leader nor strong environmental stimuli acting on the individuals.

Coordination often occurs in animal groups as the result of some kind of imitation in a broad sense, that is, when individuals take similar decisions than other group members. In the simplest scenarios, no specific social links need to be specified and a few simple assumptions can be made regarding the cognitive abilities of the organisms. For instance, in ant colonies efficient foraging decisions are taken collectively while each individual follows pheromone trails left by others (Deneubourg and Goss 1989, Bonabeau et al. 1997). On the one hand, trails are reinforced by the attraction exerted by pheromone and, on the other hand, pheromone evaporation prevents the emergence of useless paths, such as those leading to areas without resources or that are unnecessarily long. Likewise, the analysis of three-dimensional coordinated movements of starling flocks (Figure 1) reconstructed from video recordings reveals relatively simple inter-individual interactions (Ballerini et al. 2008, Bialek et al. 2014). As shown by the idealised model proposed by Viscek et al. (1995), moving in the same direction can be achieved in very large groups provided that each individual tends to orient its velocity along the average flight direction of its immediate neighbours. This alignment rule can also be made stochastic by introducing some random angular noise in the orientation decisions, which may represent either the inability of each individual to estimate exactly the heading directions of neighbours, or some type of "free-will". If this individual noise exceeds a critical threshold, or if the density of neighbours becomes too small, collective motion can disappear abruptly and be replaced by disordered states composed of non-coordinated individual movements. The larger the total number of individuals in the group, the more abrupt is the transition (Viscek et al. 1995). In order to strengthen spatial cohesion, Couzin et al. (2002) introduced an agent-based model that incorporates attraction as well as short-range repulsion, in addition to the alignment rule. Depending on the interaction parameters, this model displays rich dynamical behaviours that resemble those observed in real groups, such as swarms, torus, or dynamical parallel groups (Couzin et al. 2002).

Real flocks are not only an illustration of coordinated, cohesive movement taking place in groups, they also exhibit complex internal dynamics that allow a fast transfer of information among individuals, without loss and keeping the variance of individual velocities small (Bialek et al. 2014). Groups can respond quickly to external perturbations (for instance, the presence of a predator), even if only a small fraction of individuals are actually informed of such perturbations. Using the type of individual movement rules described above, simulations show that a small number of informed individuals is sufficient to accurately orient the whole group towards a preferred direction and with a low probability of fragmentation (Couzin et al. 2005). This type of information transfer does not require signalling or that group members know which individuals have information. In addition, the presence of long-range spatial correlations between individual velocities, observed in experiments with starlings, suggests that these systems are poised to respond maximally to environmental perturbations (Cavagna et al. 2010).

The cohesion of large groups can fluctuate in time, as individuals interact with their environment. Sheep herds (Ginelli et al. 2015) or troops of baboons (Strandburg-Peshkin et al. 2015), alternate slow foraging phases during which they spread out, with fast aggregation events that maintain cohesion and are triggered by the movements of initiators. Field observations in sheep can be explained by an individual based-model inspired from Couzin et al. (2005) that includes resting, walking and running phases (Ginelli et al. 2015), and where the attractive force increases sharply only when the distance between two neighbours exceeds a certain value. The individual rules generate a collective intermittent dynamics of fast aggregation events spaced in time, which possibly resolves the dilemma of having enough foraging space to avoid inter-individual competition, combined with the imperative of protection against predators offered by cohesion. Despite the fact that they belong to more socially stratified societies, baboons on the move exhibit similar spatial dynamics, and, like sheep, are prone to follow democratically the largest subgroup of initiators at any time during their foraging trips, rather than dominant individuals (Strandburg-Peshkin et al. 2015). Similarly, Dottie et al (2016) showed that a group of baboons traversing habitat with different kinds of resource distributions modified its shape adaptively for efficient foraging, while still constrained by the predation risk perceived by the individuals. These shape changes seem to be a partial outcome of individuals modifying their traveling trajectory, with those in the core of the group having more influence on the positions of those who lie in the periphery of the group than viceversa (Bonnell et al. 2017).

An idea suggested from the studies above is that groups can behave as entities equipped with some "collective cognition" that emerges from interactions and that can sense the environment. This principle is not restricted to animal behaviour and is reminiscent, in a fairly different context, of the collective guidance of cells. Evidence actually shows that neural crest cells can aggregate and follow very weak chemical gradients cohesively, under conditions when single cells cannot sense gradients (Camley et al. 2016).

Spatially explicit simulation models can be useful to further understand the emergence of social organisations, for instance, social differentiation and task division. In the model of Hemelrijk (1999, 2013), a set of individuals form a group and compete by pairs. After winning a dominance interaction, an individual chases away the other. All individuals start with the same probability of winning. After a fight, the dominance parameter of the winner increases and that of the loser decreases, affecting their probability of winning the next fight. If fights are frequent, the model produces a spatial segregation of individuals according to their social rank, with the dominant individuals occupying the centre of the group, and the subordinates the periphery (acting as shields against eventual predators; as was observed in the baboon study by Bonnell et al. 2017 cited above). In addition, this model exhibits a phase transition from equalitarian to despotic organisation, as a parameter representing the propensity to fight is varied. In this view, social organisation is an emergent collective phenomenon.

A similar model based on reinforcement of past dominance interactions can explain task division in bumblebee nests (Hogeweg and Hesper 1983). If the simulated nest is divided in a central part and a periphery (as in real nests), the dominance value of an agent will affect its position in space and therefore its activity. Namely, peripheral low-ranked individuals will be more likely to forage and high-ranked ones to feed the brood. Therefore, stable specialised classes can emerge from the individual experiences of initially identical agents, hence without the need of assuming that such classes reflect genetic differences. Similar principles may be applicable to primate societies: in chacma baboons, for instance, under predation risk individual placements are determined by competitive abilities, with the dominant animals occupying the center of the group (Dostie et al. 2016).

As mentioned earlier, environmental factors are key to group formation, cohesion and structure. In particular, grouping patterns in primates and other animals can occur as side effects of feeding on the same food patches, rather than from direct attractive interactions (Sugardjito et al. 1987, Mitani et al. 1991, te Boekhorst & Hogeweg 1994). Spatially explicit agent-based models are useful to study the effects of heterogeneity in resource distribution on the formation of temporary aggregations in a population of foraging individuals (Wilson and Richards 2000; Getz and Saltz 2008). Aggregation mediated by resources is one of the mechanisms that could lead to high levels of fission-fusion dynamics in primate societies, a property that has been interpreted as socially complex and thus as a selective pressure for cognitive abilities (Aureli et al 2008). Based on observations of spider monkeys, Ramos-Fernandez et al. (2006) developed a simple foraging model where non-interacting individuals visit fruit trees of varying sizes, following mental maps and a maximum efficiency principle. The individual foraging rules can generate frequent fusion and fission events that resemble the association patterns observed in spider monkeys, in particular for intermediate levels of environmental heterogeneity. In this regime, the model predicts that the average subgroup size reaches a maximum. A network analysis is also carried out, where the nodes represent agents and the edges connect pairs of individuals that have been in contact at least once (occupying the same tree) during the simulation period. In heterogeneous environments, most of the individuals of the group belong to the same "giant" cluster of interconnected individuals. This is due to the fact that many individuals may feed at the same time during their trajectory on one of the few trees that are very rich in fruits. If the edges are weighted in proportion to the time spent together by the nodes they connect, further analysis shows that the giant cluster is itself composed of small "cliques", *i.e.*, sub-structures composed of individuals that are more densely connected to each other than to others in the group. In comparison, very scarce resource distributions induce much fewer associations, resulting in fragmented networks composed of small isolated clusters. Very abundant resources have a similar effect, as individuals may cluster around different large trees. This model shows that a complex grouping pattern can emerge from relatively simple rules guiding individual foraging decisions, without necessarily involving explicit rules modifying grouping tendencies or dyadic interactions, and that social structures are sensitive to the spatial distribution of resources.

Strong effects of environmental factors on association networks have been reported using empirical data. A network analysis performed by Mokross et al. (2014) quantified how the structure of mixed-species bird flocks were affected by habitat modifications in the Amazonian rain forest. It was found that habitat type had a strong effect on the structure of the networks of non-trophic interactions among bird species, as well as on behavioural interactions within flocks. Frequency of associations among species declined with increasing levels of forest fragmentation due to human activities. Flock cohesiveness and the density of cliques were also positively correlated with the mean vegetation height. These results suggest that social structures are comparatively more sensitive to environmental changes than other ecological networks.

**3.2 Emerging structures in network models**

The above studies illustrate the convenience of network analysis for describing systems composed of individuals that are not necessarily cohesive and have different activities taking place simultaneously. We further review a few representative network models in non-spatial contexts, as they give useful insights on the relationships between individual connectivity and emerging macroscopic properties. Precise definitions of the network measures mentioned here can be found in Newman (2003).

One of the most studied collective phenomena in networks is percolation. The percolation property is characterised by the presence of a giant cluster connecting a large fraction of a set of nodes (Bollobás and Riordan 2006). In the presence of a giant cluster, it is likely that a path connecting a pair of nodes chosen at random exists. This property is well understood in random networks (Figure 2a), which are constructed by adding connections one by one and at random among a set of initially unconnected nodes. Although random, this process leads to the emergence of a giant component when the number of connections per individual reaches a critical value (Newman 2003).

Generally, complex networks are characterised by heterogeneous connections, namely, the number of connections (or the degree) of a node can vary largely from one node to the other (Figure 2b). The presence of a few highly connected connected nodes, or "hubs", characterised by a degree significantly larger than the average degree of the network, is very rare in random networks but commonly observed in technological, biological or social networks (Barabási and Albert 1999). In animal social networks, node heterogeneity can play an important role in information flow (Lusseau and Newman 2004, Pinter-Wollman et al. 2014).

Since random networks have homogeneous (Poissonian) degree distributions, other rules must be implemented to account for the observed heterogeneity of many real world networks. For instance, networks can be grown by adding nodes one by one and by connecting them to previously existing nodes (Krapivsky et al. 2000). If each new node is connected to a randomly chosen existing node, a homogeneous network with exponential degree distribution is obtained. The preferential attachment rule, in turn, consists in connecting the new node to an existing node which is chosen with a probability proportional to its degree. Therefore, highly connected nodes tend to attract the new connections. This process produces non-exponential networks, characterised instead by power-law degree distributions that entail the presence of highly connected nodes (Barabási and Albert 1999). These "hubs" are usually the older ones, and the network as a whole grows hierarchically, as the degree of a node is highly correlated with its order of appearance. Primate social networks do not seem to be scale-free, but those of other species like bottlenose dolphins or Columbian ground squirrels exhibit such structure (see Kanngiesser et al. 2011 for a discussion).

A peculiarity of many social networks, compared to other types of networks, is their cliquishness (Newman and Park 2003). The clustering coefficient is the likelihood that two nodes connected to a same node are themselves connected to each other. When connections are transitive, this coefficient is high (Figure 2c-d). In random networks or networks constructed according to preferential attachment, cliques are practically absent and the clustering coefficient very small. Other models that generate dense networks have helped to understand how cliquishness can confer resilience to a system. In the model of Marsili et al. (2004), a network is constantly perturbed by random deletion of links. In parallel, the nodes seek to establish new connections either randomly in the population or via "friends". In the latter case, a node can connect to a neighbour of one of its own neighbours. The dynamics of this model can be viewed as a continuous struggle against volatility (link deletion) with connections arising from new opportunities partially mediated by the existing network. In an interval of values for the rate of the connections via friends, the resulting networks are both densely connected and with a very high clustering coefficient. Surprisingly, these structures are not the unique solutions, since, depending on the initial conditions, networks with low average degree and low clustering coefficient can also form for the same parameter values. These results imply that dense networks, in which short range reconnections are allowed, can be robust with respect to node deletion and may even resist under conditions in which dense networks would not form.

A commonly studied local property, giving information on the position of a node in its network, is centrality. The betweenness centrality of a node $i$ is the number of shortest paths joining other nodes (geodesics) that run through $i$. Central nodes are not necessarily the most connected ones, but their removal may considerably increase the length of the shortest path for going from one node to another (see Newman 2003 and refs. therein). Eigenvalue centrality is defined from the eigenvalue spectrum of the network adjacency matrix, and related to diffusion processes in the network. These two measures are similar and tell about network resilience, namely, the effect that node removal will have on path lengths and therefore network topology and function.

**3.3 From structure to interactions: functions of complex networks**

Simulation models are useful to illustrate how macroscopic structures can be the outcomes of individual behaviours and to predict un-observed but possible features by varying parameters. Note that in the vicinity of a phase transition, mild changes in behaviours produce drastic differences at larger scales. Conversely, as shown below, macroscopic features crucially contribute to orient individual behaviour, even when agents are not aware of them. One may then view social dynamics as a constant interplay between the evolution in time of large scale structures and that of individuals with local (e.g. dyadic level) interactions. A ongoing challenge in social network analysis is to understand how a given architecture may facilitate certain social behaviours or certain types of dynamics.

Robustness is a property of many complex systems and certain network architectures are known to be robust under attacks. Structural robustness can be probed theoretically by node (or link) deletion. If nodes are removed at random in a scale free-network, for instance, the mean topological distance between any pair of nodes remains practically unaffected, in sharp contrast with the case of homogeneous networks (Albert et al. 2000, Newman 2006). In this sense, scale-free networks are robust. However, the situation is quite different if attacks are not random but targeted, namely, if the most connected nodes are progressively deleted. In this case, the giant cluster component does not remain cohesive and the mean topological distance betweens pairs of nodes increases sharply with the fraction of deleted nodes. The removal of very few nodes may even fragment a network into two or more components, thus preventing communication in the whole system. In empirical networks, these strategic nodes, or "brokers", are not always the most connected actually, but they are often located at the boundary between modules or communities (Newman 2006). These nodes have a high betweenness centrality, *i.e.*, they frequently lie on the shortest path that connects a pair of nodes randomly chosen in the network. By using centrality measures, Lusseau and Newman (2004) have identified brokers in the association network of a group of bottlenose dolphins. The removal of the most central individual was predicted to substantially affect the mean distance between pairs of individuals in the network. A fission of the group was actually observed when this individual happened to temporarily leave the group. Similarly, the resilience of the grooming network of a captive chimpanzee group was probed by removing central individuals, highlighting their key role on group cohesiveness (Kanngiesser et al. 2011).

Centrality measures can thus help to understand the role that individuals play in a social network, and how collective decisions are taken. King et al. (2011) showed with the help of a network analysis that each departure from the sleeping site in chacma baboons was a self-organising process, initiated by an individual that was followed by the rest of the group. The initiators tended to be more sociable individuals, namely, those that were central in the grooming and spatial association networks, rather than dominant individuals. These findings adds support to the idea that collective behaviour in animal groups can follow relatively simple cognitive heuristics. Brent et al. (2015) further provided evidence that during the collective foraging trips of killer whales, leadership was taken by the older, more experienced individuals, specially in difficult times of low prey abundance. Such leadership also provides a simple mechanism by which ecological knowledge can be transferred within the group. Whether being a broker or a leader requires more sophisticated cognitive abilities remains elusive.

The structure of a social network can also be important for the way in which collective decision-making (and therefore group cognition) can occur. Dense networks, or networks with long-distance connections or with particular individuals functioning as hubs could be more efficient in exchanging information about the environment and responding appropriately. For example, in an experiment with human subjects, networks of cooperators that were structured in spatially based cliques were more efficient in solving problems that required broad exploration, whereas networks with greater long-range connectivity were better at solving problems that required less exploration (Mason et al. 2008).

From the above examples, one sees that networks are not just static objects made of nodes and links, but also structures on which many dynamical processes can take place. Two well-studied processes are epidemic spreading (Pastor-Satorras et al. 2015) and synchronisation of activities (Arenas et al 2008). Understanding how the structure of a network determines the evolution of its elements is of primary importance. In this context, resilience, a key feature of complex systems, refers to the ability of a system to readjust its activity in response of environmental or internal changes. In principle, it is determined both by the structure of the network and the dynamics that describe the evolution of the states of the nodes. Inspired by ecological and regulatory systems, Gao et al. (2016) have considered a general framework where a set of coupled dynamical elements are located on a fixed network of interactions. The study of the stationary states and their stability shows that global states of high activity can undergo sudden transitions to undesired states of much lower activity, when environmental conditions or interaction parameters are changed. A system is said resilient if the states of high activity remain stable over a broad range of parameter values. Networks that are dense and heterogeneous are found to be more resilient (Gao et al. 2016).

The resilience of animal social networks undergoing natural "knock-outs" can be studied in the field. Barrett et al. (2012) observed the effects of the death of dominant females in a group of baboons. The resulting instability in the social hierarchy was counterbalanced by re-structuring into more cliquish groups. The results suggest that adjustments are aimed at reducing the uncertainty of relationships, with no need for specific mechanisms of conflict management (Barrett et al. 2012; see also Flack et al. 2006 and section 4 below).

As mentioned earlier, a property of social networks is the presence of sub-communities or modules (Newman 2006). One might expect that any social network would respond to a mild perturbation by re-adjustments that preserve modularity. Modularity is a particularly important network property, as it can orient individual behaviour toward cooperation. Marcoux and Lusseau (2013) simulated prisoner's dilemma games on static networks of varying modularity index and found that modular networks significantly promoted the evolution of cooperation (Nowak and Sigmund 1998). Therefore, cooperation could evolve in social groups without complicated mechanisms such as genetic similarity, provided that the underlying social networks were sufficiently modular. This conclusion has to be contrasted with Darwinian theories, that predict the evolution of selfish individual when interactions take place randomly in a population. In fact, in an analysis of 70 interaction networks of 30 different primate species, Voelkl and Kasper (2009) found that the structure was precisely what favoured the fixation of cooperative traits, compared to well-mixed populations of similar size or randomly connected networks. Similarly, cooperation happens to be the dominating trait in evolutionary games that take place on scale-free networks (Santos and Pacheco 2005). This latter result differs completely from the outcomes of binary games or games on regular networks, where under the same rules, cooperators are unable to resist the invasion by defectors.

## 4. Can social network metrics be used as an indicator of social and cognitive complexity?

It is tempting to use metrics derived from social network analysis as measures of social complexity. The density of links, the heterogeneity in the degree distribution or the modularity of a network could be considered as measures of its complexity (Costa et al. 2007). In keeping with the most commonly used definition of social complexity (Dunbar 1988), one could use the number of different interactions (e.g. grooming, proximity, co-feeding, etc.) that would need to be included in a multi-layered network that effectively described the social structure of a particular species (Barrett et al. 2012).

However, our review shows that many structural and functional properties of groups and their social networks are an emerging result of individual interactions, without necessarily implying that the complexity at the collective level rests on complex cognitive processes. If we do not know which aspects of social networks are actually taken into account by individuals in their decision-making, it is not clear which metrics of social network analysis could be useful for defining social complexity in a way that is relevant for the study of social cognition.

Brent (2015) argues that while primates do seem capable of understanding third party relationships, it is not known whether they consider other, more distant individuals in their social network, to whom they are connected only indirectly, in their decision-making. An intriguing aspect of real-world networks is the existence of recurring substructures or "motifs" (Milo et al. 2002). These appear to be the result of constraints during the evolution of a network and would function as the elementary building blocks or computational circuits through which information flow and other processes occur in a network. The assembly of these relatively simple building blocks would naturally lead to large functional structures such as social networks. It is intriguing to think that social interactions also lead to the occurrence of these motifs (such as triadic closure in affiliative relationships or alliance formation in agonistic contexts) and that the whole network would grow as a result of these local processes coming together. Moreover, in terms of cognition, motifs could be recognised by individuals using relatively simple pattern-recognition mechanisms, which would also allow them to act in appropriate ways without necessarily taking the whole network into account (Barrett et al. 2007).

In the case of network growth, as we review in section 3.2, a process that can lead to a heterogenous (and therefore complex) distribution of network degrees is preferential attachment, i.e. new nodes joining those nodes that are already well-connected. If animals recognise these highly connected nodes when establishing their social relationships, it would imply that they have some way of recognising the connections that go to others beyond their own connections, or at least some proxies for the highly connected nodes.

Several studies have shown that animals can modify the structure of their network in predictable ways. For example, the ground-breaking study of Flack et al. (2006) showed experimentally that particular individual macaques with "policing" roles in their social network can affect the structure of the network as a whole. Removing these particular individuals resulted in a change in all interactions, producing less diverse and more fragmented networks. Clearly, from a purely functional standpoint, policing stabilises and promotes cohesiveness within the social network. However, we know little about the cognitive mechanisms underlying this policing role or the effect that the absence of a policer has on the decision-making by others.

Building on Flack et al. (2006), Barrett et al (2012) have proposed that social animals develop multi-layered interaction networks precisely in order to reduce the uncertainty about their social relationships (Figure 3). In other words, more interaction contexts (and thus more social complexity) reduce uncertainty in predicting the behaviour of social companions. A perturbation of the network in one context (as in the knockout experiments by Flack et al. 2006 or the natural observations by Barrett et al. 2012) leads to adjustments in other contexts, which reduce the uncertainty (measured by a decrease in Shannon's entropy) of the whole network. This supports the idea that a multi-layered social network is the best representation of social structure and that different layers (or interactions) are interrelated. It remains to be shown to what extent are animals actively making these adjustments as a result of their understanding of their social network.

**Acknowledgements**

We acknowledge financial support from DGPA-PAPIIT grant IN105015, CONACYT grant 157656 and Instituto Politecnico Nacional. We thank Louise Barrett for fruitful comments on the manuscript.

**Figure legends**

*Figure 1: Starling murmuration at the Royal Society for the Protection of Birds (RSPB) nature reserve at Minsmere, Suffolk, UK. Picture by Airwolfhound, Creative Commons License CC BY-SA 2.0.*

*Figure 2: Types of network architecture. a) A connected component with 40 nodes and 39 edges extracted from a random network, where every pair of nodes has the same probability of being linked. The degree distribution (or the number of connections per node) would be Poissonian; b) A network of 40 nodes and 39 edges, established with preferential attachment, i.e. nodes with more edges have a higher probability of having more edges and becoming "hubs". The degree distribution of such a network would be non-Poissonian (and for very large networks, scale-free); c) A network with high modularity or cliquishness, where 40 nodes are clustered into four components (which themselves are random networks with a probability of an intra-module edge being 0.8 and only 5 edges joining each module to others). A clear separation into four modules can be observed and some nodes linking different modules would be*

*relevant as "brokers"; d) A network with low modularity, composed of the same number of nodes and edges as the network in c) but with a probability of 0.66 of an intra-module edge and 10 edges joining modules. No clear separation into modules can be observed.*

*Figure 3. A conceptual model of social structure by means of a multi-layered network, with each layer corresponding to a different interaction and the whole social network being the object containing these different layers. Modified from Barrett et al 2012.*

Figure 1

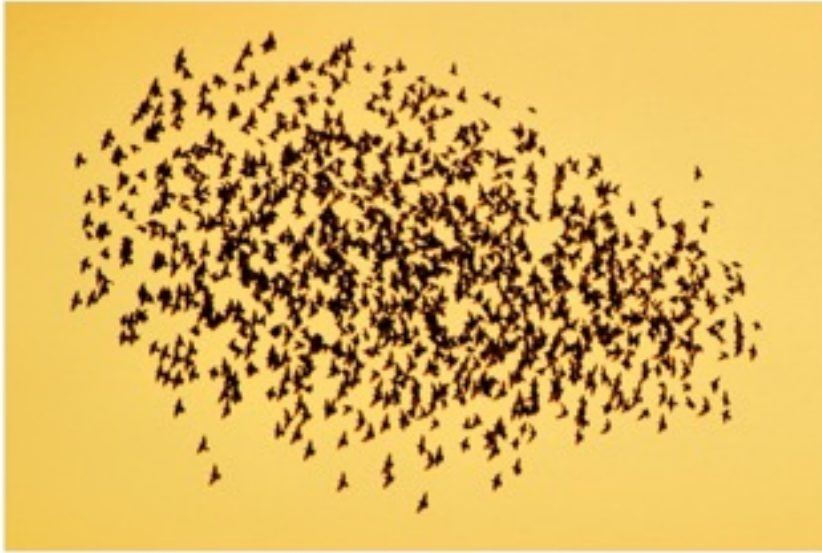

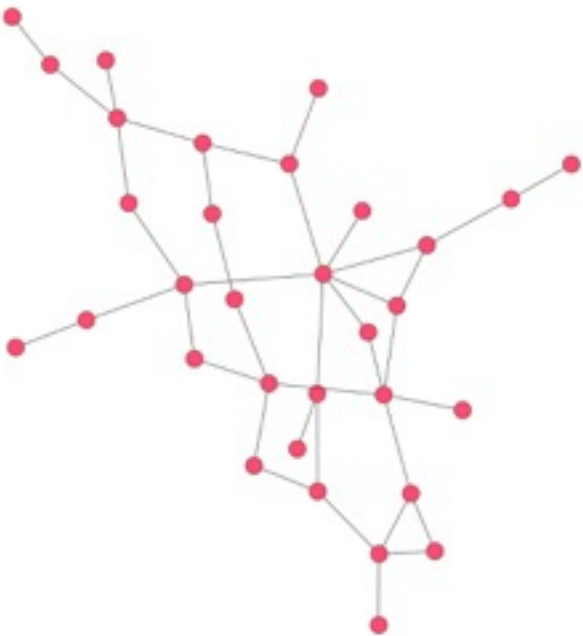

Figure 2a

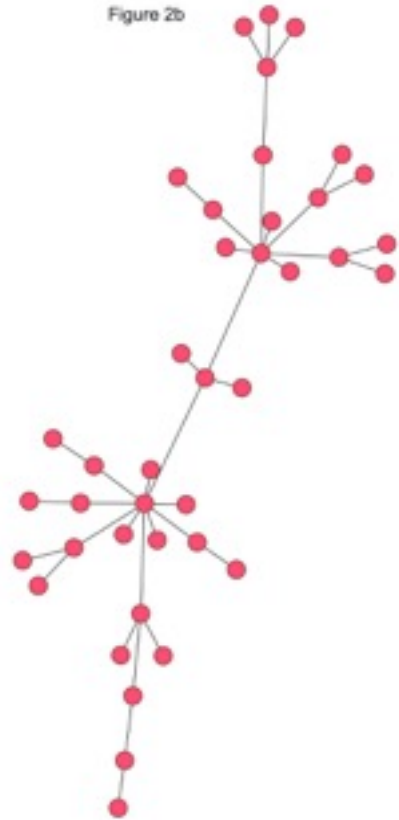

Figure 2b

Figure 2c

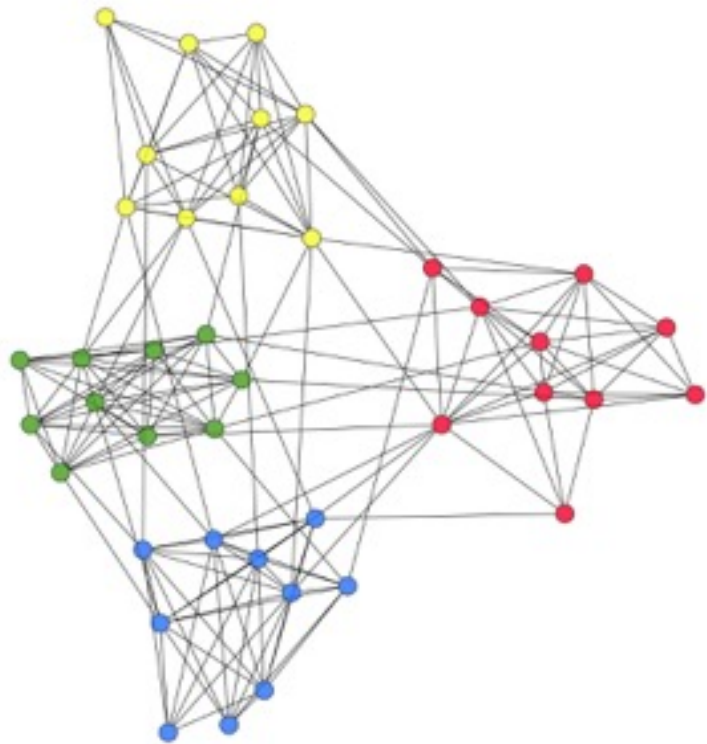

Figure 2d

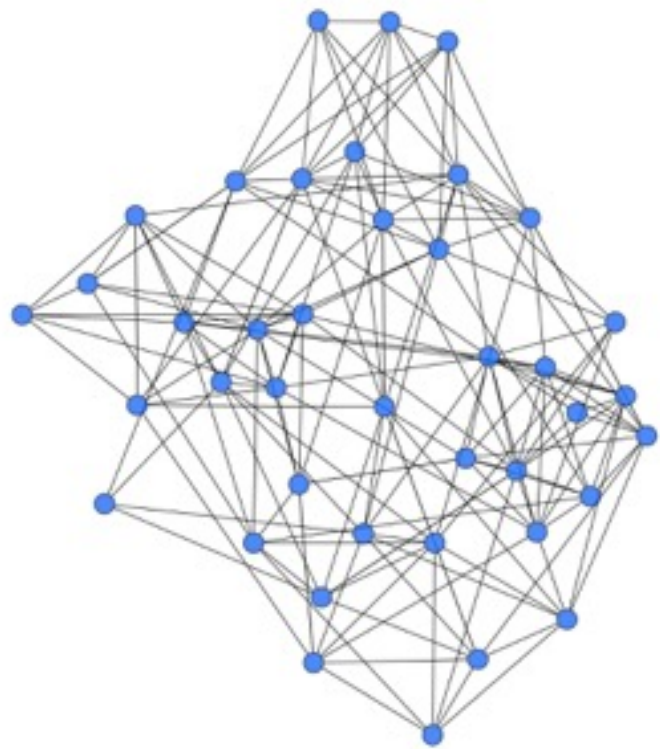

Figure 3

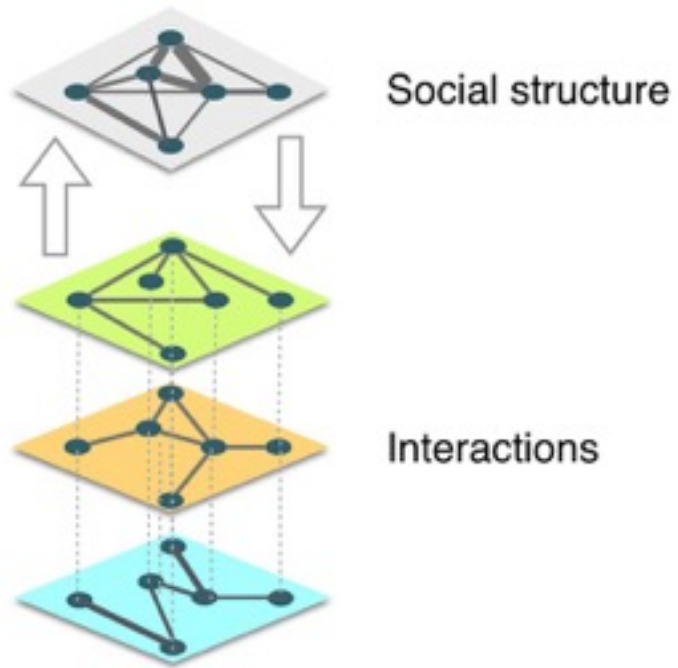